\documentclass[letter]{aa}  

\usepackage{xcolor}
\usepackage{hyperref}
\usepackage{graphicx}
\usepackage{txfonts}
\usepackage{lipsum}
\usepackage{subcaption}         % necessary for continued figures, example in section 3
\usepackage{lscape}             % to rotate a single page table, example in appendix.
\usepackage{placeins}           % useful with \FloatBarrier, to keep 
\usepackage{lineno}
\linenumbers
\renewcommand\thelinenumber{}
                                
\newcommand{\dpn}[3]{$#1\,{\rm #2}_{#3}$}
\newcommand{\ccc}[1]{#1} %modificaciones en negro

\defcitealias{Munoz26}{Paper I}

\begin{document}

%%%%%%%%%%%%%%%%%%%%%%%%%%%%%%%%%%%%%%%%
% if you use custom commands in your title,
% ensure to check your title when submitting!
%%%%%%%%%%%%%%%%%%%%%%%%%%%%%%%%%%%%%%%%

   \title{On the Influence of Pluto on Twotino Dynamics Through Their Mutual 4:3 Mean Motion Resonance}
    
%%%%%%%%%%%%%%%%%%%%%%%%%%%%%%%%%%%%%%%%
% Please separate each author with the \and command
%
% Please do not include ORCIDs next to author names.
% Only ORCIDs authenticated by individual authors in EDPS
% editorial system will be taken into account.
% ORCIDs included here will be removed.
%%%%%%%%%%%%%%%%%%%%%%%%%%%%%%%%%%%%%%%%

   \author{S. Ram\'irez-Vargas\inst{1}
        \and A. Peimbert\inst{2}
        \and M. A. Mu\~noz-Guti\'errez\inst{3}\fnmsep\thanks{Corresponding author: marco.munoz@uda.cl}
        \and A. Perez-Villegas\inst{4}
        }

   \institute{Instituto de Astrof\'isica, Pontificia Universidad Cat\'olica de Chile, Av. Vicu\~na Mackenna 4860, 782-0436 Macul, Santiago, Chile
   \and Instituto de Astronom\'ia, Universidad Nacional Aut\'onoma de M\'exico, Apdo. postal 70-264, Ciudad Universitaria, M\'exico
   \and Instituto de Astronom\'ia y Ciencias Planetarias, Universidad de Atacama, Copayapu 485, Copiap\'o, Chile
   \and Instituto de Astronom\'ia, Universidad Nacional Aut\'onoma de M\'exico, A. P. 106, C.P. 22800, Ensenada, B.C., M\'exico}
%   \and Department of Astronomy, Indiana University, Bloomington, IN 47405, USA}

   \date{Received \today}

% \abstract{o}{m}{m}{m}{o}
% 5 {} token are mandatory
 
\abstract
  % context heading (optional)
   {The role of Pluto contributing to the long-term evolution of the trans-Neptunian region has been considered significant only over its neighboring Plutinos. However, it has recently been found that the long-term stability of the Twotino population is strongly affected when including Pluto as a massive object in simulations, while Eris, with a similar mass, has a negligible effect.}
  % aims heading (mandatory)
   {We hypothesize that the effect of Pluto on Twotinos results from the latter being trapped in a 4:3 mean motion resonance (MMR) with Pluto. In this work, we aim to demonstrate the resonant behavior of Twotinos within Pluto's 4:3 MMR and the significance of this resonance for the long-term evolution of the population.} 
  % methods heading (mandatory)
   {We run high-resolution, 10 Myr REBOUND simulations of the observed Twotino population in the Kuiper belt, under the perturbations by the Sun, the four giant planets, and Pluto, as massive objects.}
  % results heading (mandatory)
   {We find that all objects trapped in the 2:1 MMR with Neptune are locked in a weak 4:3 MMR with Pluto. The 4:3 resonant argument of most objects trapped in the leading and trailing islands of the 2:1 MMR, librates with amplitudes lower than $360^\circ$. Objects in the symmetric islands of the 2:1 MMR librate in the 4:3 MMR with amplitudes greater than $360^\circ$, but, contrary to circulating objects, will oscillate by up to $850^\circ$ visiting preferred regions on Pluto's co-rotating frame, indicating a diluted resonant effect that may also perturb their orbits on secular timescales.}
  % conclusions heading (optional), leave it empty if necessary
   {The importance of Pluto in shaping the structure of the trans-Neptunian region should be reconsidered, \ccc{specifically} for resonant populations. Moreover, with current computational power, its exclusion from simulations cannot be justified.}

   \keywords{Methods: numerical --
                Kuiper belt: general --
                Kuiper belt objects: individual: Pluto
               }

   \maketitle

%%%%%%%%%%%%%%%%%%%%%%%%%%%%%%%%%%%%%%%%%%%%%%%%%%%%%%%%%%%%%%

% JUSTIFICATION LETTER

%Pluto's MMRs have never been considered as a possible ingredient in the evolution of the TN region. However, given the 3:2 MMR between Pluto and Neptune, any object in Neptune's MMRs has a chance to be trapped in a MMR with Pluto. In this work, we show this to be the case for Twotinos; being in 2:1 MMR with Neptune puts them in a 4:3 MMR with Pluto, and this resonance may contain the mechanism through which Pluto is capable of significantly affecting their long-term evolution, as found in our previous work (Muñoz-Gutiérrez et al. 2026). We consider this result to be highly significant for the community, besides the fact that it has not been recognized neither taking into account in the past. For these reasons, we deem this idea to deserve an expedited publication as a Letter. 

\section{Introduction}

Long after Clyde Tombaugh's discovery of Pluto in 1930, the first long-term integration of its orbit \citep{Cohen65}, aided by the first computers, enabled the determination of its resonant state with Neptune. Some of the pioneering works that studied the secular dynamics of the outer Solar System, from the late 1980s to the early 1990s, considered the perturbations of the Sun and the four giant planets, together with an ever larger number of test particles \citep[see, e.g.][]{Duncan88,Torbett89,Gladman90,Levison93}. It probably seemed wasteful to include other massive bodies, since Uranus is over 6,500 times more massive than Pluto\footnote{See e.g., \url{https://ssd.jpl.nasa.gov/planets/phys_par.html}, and references therein.}, and over 5,200 times more massive than Eris, the two most massive currently known trans-Neptunian objects (TNO).

%In the 1990s, with larger computers, it became feasible to study the TNR using larger numbers of massive bodies; however, they have typically been restricted to the Sun, the four giant planets, and larger and larger numbers of test particles. It probably seemed wasteful to include other massive bodies, since Uranus is over 5000 times heavier than Eris, the heaviest currently known TNO. 

It has also been recognized that Pluto has an important influence over objects in its vicinity, being responsible, among other things, for the destabilization of other Plutinos, which contribute to the injection rate of comets into the inner Solar System \citep{Morbidelli97,Yu99,Nesvorny00,Tiscareno09,Munoz19}. 

\ccc{However, Pluto can have a secular effect on other objects besides its neighboring Plutinos. In \citet[][hereafter \citetalias{Munoz26}]{Munoz26}, we found that Pluto has a significant effect on the leaking rate of the 2:1 mean motion resonance (MMR); especifically, the stable fraction of Twotinos after 4 Gyr of evolution is reduced from 47\% when considering only the giant planets, to 19\% when including Pluto along the giant planets as a massive object in simulations.}

%However, Pluto can have a secular effect on other objects besides its neighboring Plutinos. In particular, Twotinos in 2:1 mean motion resonance (MMR) with Neptune appear to be directly affected by Pluto, exhibiting a significantly larger leaking rate from the resonance when Pluto's mass is considered compared to when it is ignored \citep{Munoz26}.

\ccc{The origin of this significant effect is not straightforwardly related to Pluto's mass. In \citetalias{Munoz26}, we found that Eris with a similar mass, produces an effect an order of magnitude smaller than Pluto on the Twotinos. We note, however, that although Pluto is not massive enough to capture other TNOs into its own MMRs, Neptune has captured enough objects in its resonances, and those will therefore be at the same time in resonance with Pluto.} 

\ccc{In this letter, we show the geometrical configuration between Pluto and Twotinos corresponds to a previously unrecognized 4:3 MMR between them, and that the cumulative resonant effects from Pluto will slowly but surely distort Twotinos' orbits. This shows that Pluto is an important ingredient when studying the secular evolution of the outer Solar System, and with current computational capabilities, its exclusion from simulations cannot be justified.}

%Although Pluto is not massive enough to capture other TNOs into its own MMRs, Neptune has captured enough objects in its resonances, and those will therefore be at the same time in resonance with Pluto. In this letter, we will show that Pluto's effect on Twotinos may be due to a previously unrecognized 4:3 MMR between them. 

%This makes Pluto an important ingredient when studying the secular evolution of the outer Solar System, and with current computational capabilities, its exclusion from simulations cannot be justified.

%This letter is organized as follows. Section \ref{Sec:Connnetion} presents the connection of Pluto and Twotinos with Neptune. Section \ref{Sec:Validation} provides the validation of the 4:3 MMR in numerical simulations. Section \ref{Sec:DynamicalSignificance} discusses the dynamical significance of the 4:3 resonance and the effect of Eris. Finally, Section \ref{Sec:Conclusions} presents our conclusions.

\begin{figure*}
    \centering
    \includegraphics[width=1\linewidth]{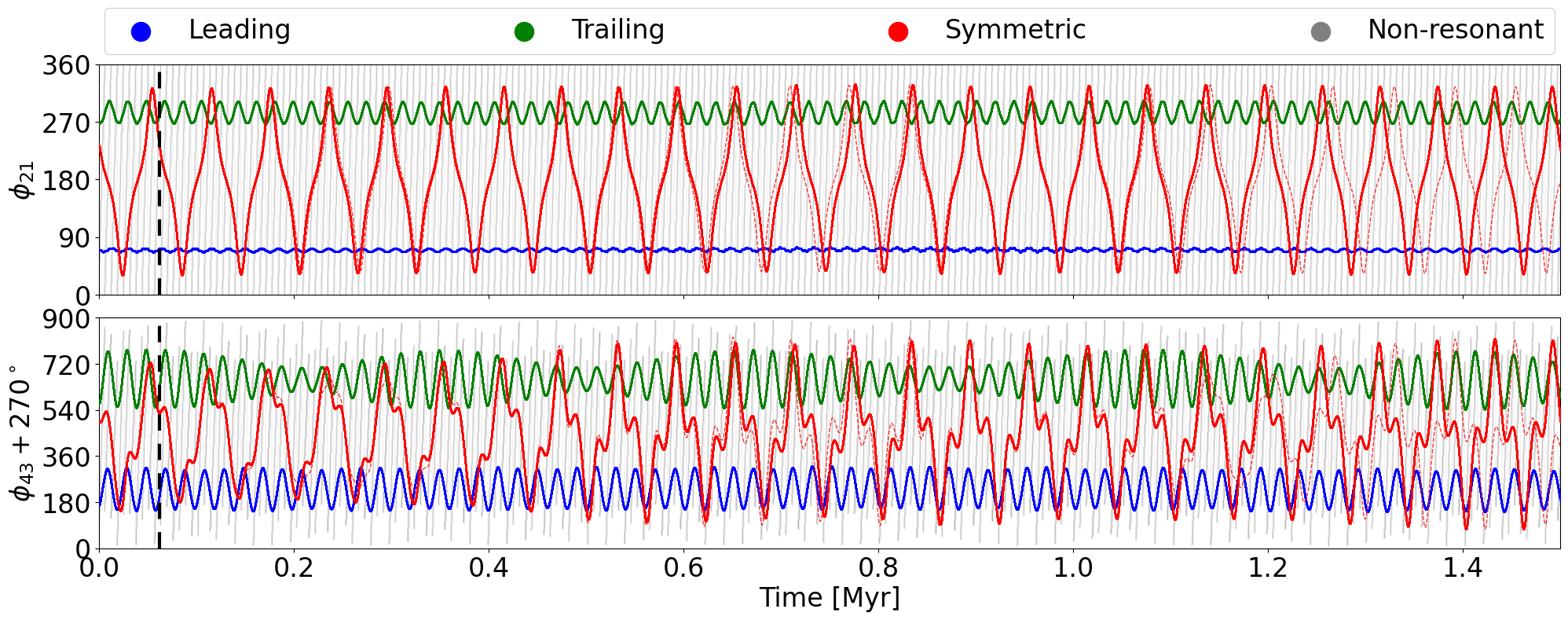}
    \caption{Evolution of the 2:1 resonant arguments with respect to Neptune (top panel) and the 4:3 resonant arguments with respect to Pluto (bottom panel) for four selected objects. The blue (\dpn{2015}{VH}{166}, leading), green (\dpn{2015}{GZ}{54}, trailing), and red (\dpn{2015}{KX}{173}, symmetric) curves show the behavior in the three possible Twotino libration islands. The gray curve shows a circulating (\dpn{1999}{RX}{215}, non-resonant) object. \ccc{Each Twotino has a corresponding dashed line, indicating the evolution of the same object, but without the gravitational influence of Pluto; at this timescale, differences between the two scenarios are only noticeable for the Twotino in the symmetric island.} The vertical dashed line is placed at $62,000$ yr, representing one libration period of the symmetric object.}
    \label{fig:phitime}
\end{figure*}

\section{The Connection Through Neptune}\label{Sec:Connnetion}

\ccc{As both Pluto and any Twotino are locked in MMR with Neptune, they are also indirectly locked in resonance with each other.} While Neptune orbits six times around the Sun, Pluto will orbit four times, and a Twotino will orbit three times. Therefore, Twotinos will be in a 4:3 MMR with Pluto.

%Pluto is locked in a 3:2 MMR with Neptune, and by definition, a Twotino is locked in a 2:1 MMR with Neptune. As they are both locked in resonance with Neptune, indirectly, they are also locked in resonance with each other. While Neptune orbits six times around the Sun, Pluto will orbit four times, and a Twotino will orbit three times. Therefore, Twotinos will be in a 4:3 MMR with Pluto.

\ccc{It is well known that resonant arguments of Pluto and Twotinos in Neptune's 3:2 and 2:1 MMRs are, respectively:}
\begin{eqnarray}
\label{equ:phi32}
\phi_{3:2} & = & 3 \lambda_{\rm P} - 2\lambda_{\rm N} - \varpi_{\rm P} \\
\label{equ:phi21}
\phi_{2:1} & = & 2 \lambda_{\rm T} - \lambda_{\rm N} - \varpi_{\rm T}
\end{eqnarray}
%It is well known that Pluto's resonant argument in Neptune's 3:2 MMR is given by:
%\begin{equation}\label{equ:phi32}
%\phi_{3:2} = 3 \lambda_{\rm P} - 2\lambda_{\rm N} - \varpi_{\rm P}
%\end{equation}
%while Twotinos show libration of the angle:
%\begin{equation}\label{equ:phi21}
%\phi_{2:1} = 2 \lambda_{\rm T} - \lambda_{\rm N} - \varpi_{\rm T},
%\end{equation}
where $\lambda$ is the mean longitude, $\varpi$ the longitude of perihelion, and the sub-indices P, N, and T stand for Pluto, Neptune, and Twotinos, respectively. To cancel the dependence on Neptune's mean longitude, we can subtract $\phi_{3:2}$ from twice $\phi_{2:1}$.
\begin{eqnarray}\label{equ:phi43}
\phi_{4:3} &=& 2 \phi_{2:1} - \phi_{3:2} \nonumber \\
&=& 4\lambda_{\rm T} - 3 \lambda_{\rm P} - 2 \varpi_{\rm T} + \varpi_{\rm P}.
\end{eqnarray}
The libration of this resonant argument corresponds to Twotinos being locked in a first-order, 4:3 MMR with Pluto \citep[with $\phi_{4:3}$ corresponding to eq. 4D1.4 in][]{murray99}. This resonance is weaker than the ones involving Neptune, as its relative strength includes a third-order eccentricity term ($e_Pe_T^2$), instead of first-order terms as other resonant arguments with Neptune do ($e_P$, and $e_T$, respectively); it is also slightly more complex as it involves the longitudes of perihelion of both resonant objects.

\ccc{It should be noted that the 4:3 MMR is part of a} 6:4:3 three-body resonance between Neptune:Pluto:Twotinos, but the effects of Neptune on the Twotinos and Pluto through the 2:1 (6:3) and 3:2 (6:4) MMRs have been studied extensively \citep[e.g.,][]{Williams71,Malhotra96,Gallardo98,Nesvorny01,Kotoulas05}, \ccc{while the effects of Pluto on the Twotinos have not been studied previously. Here, we focus on the relationship between Pluto and the Twotinos as a first-order 4:3 MMR, without forgetting that, given Pluto's mass being four orders of magnitude smaller than Neptune's, the effect of Pluto on the Twotinos can be considered as a perturbation to Neptune's dominating dynamics. Thus, it is justified to study an isolated two-body resonance, even if the factor that puts those objects in resonance is an external element. Overall, we stress that $N$-body simulations with a massive Pluto are required to correctly model the secular evolution of the trans-Neptunian region.}

The libration of Pluto's 3:2 resonant argument has a large amplitude \citep[from $\sim160^{\circ}$ to $\sim172^{\circ}$, e.g.][]{Milani89,Malhotra2022}. There are also relatively large librations of Twotinos in the three possible resonant islands of the 2:1 MMR, especially for Twotinos in the symmetric island \citepalias[e.g. Figure 4 of][]{Munoz26}; thus, there will be some libration between Pluto and the Twotinos; since the above librations are not correlated, the total libration of the 4:3 MMR will be large. Consequently, the total libration amplitude of $\phi_{4:3}$ will amount to $\sim170^{\circ}$, from the libration of Pluto relative to Neptune, plus twice the librating amplitude of any particular Twotino relative to Neptune. 

In principle, this would restrict the 4:3 MMR to Twotinos with librations smaller than $95^{\circ}$ (i.e. Twotinos in the leading and trailing islands); however, objects where librations of $\phi_{4:3}$ go beyond $360^{\circ}$ are not circulating in the most conventional sense, as they will be going always forward exactly once, for Twotinos whose libration relative to Neptune lies between $95^{\circ}$ and $275^{\circ}$, or twice, for Twotinos that librate between $275^{\circ}$ and $360^{\circ}$, and then they will circulate backwards the same number of turns, always being bound by up to $890^{\circ}$. \ccc{Therefore, regardless of the amplitude of libration of $\phi_{4:3}$, we will consider such particles as resonant as long as $\phi_{4:3}$ is bounded inside those limits. Despite the large libration amplitude,} the resulting resonant effect will be present (as we will prove in Section \ref{Sec:DynamicalSignificance}), though diluted, similar to the weakening present in higher-order MMRs \citep[such as the 5:3, 7:4, etc., see e.g.,][]{Tamayo25}. 

\section{4:3 MMR Validation in Numerical Simulations} \label{Sec:Validation}

\begin{figure*}[t]
    \centering
    \includegraphics[width=1\linewidth]{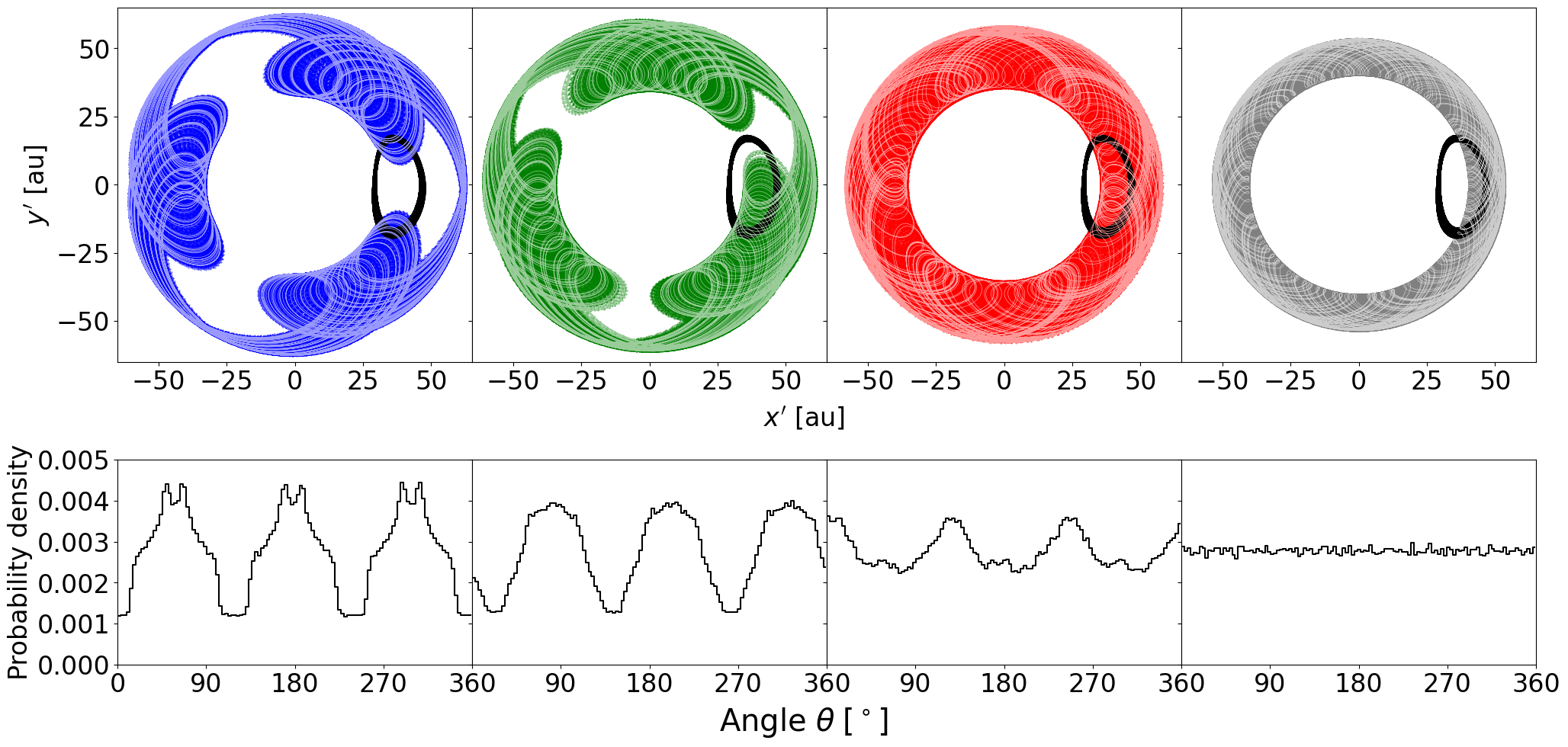}
    \caption{Co-rotational orbits with respect to Pluto of the four sample Twotinos (top panels) over \ccc{1.5 Myr}, as well as the measured probability distribution function to find each Twotino at a certain co-rotating angle (bottom panels). The black curves represent Pluto's orbit in the co-rotational frame, while colored curves represent each Twotino, following the same color scheme as Figure \ref{fig:phitime}. To see the evolution of the orbit, see the attached animation of these four objects on the \href{https://github.com/Sramrev/Pluto-Twotino-MMR.git}{GitHub} repository.}
    \label{fig:corot}
\end{figure*}

For our numerical experiments, we followed a similar procedure as in \citetalias{Munoz26}. We conducted short-term, 10 Myr simulations of the 152 Twotino candidates retrieved from the JPL's Small-Body Database \footnote{\url{https://ssd.jpl.nasa.gov/horizons/}} with a semimajor-axis in the range $47<a<48.5$ au. Our simulations include the Sun, the four giant planets, and Pluto\footnote{We consider a single body with a mass equal to the mass of Pluto and Charon together, with masses taken from \citet{Stern18}.} as massive bodies.

We perform our simulations with REBOUND \citep{rebound}, using the \ccc{IAS15 \citep{reboundias15}} integrator. We set an initial time-step of \ccc{100} days and an accuracy parameter of $10^{-9}$. We use a 5-yr output cadence to follow the evolution of the resonant orbits with high precision. Following equations \ref{equ:phi21} and \ref{equ:phi43}, we obtain the 2:1 MMR resonant argument with respect to Neptune, and the 4:3 MMR resonant argument with respect to Pluto for each Twotino. Overall, we obtain 91 resonant objects librating in the 2:1 MMR with Neptune, i.e., those with libration amplitudes \ccc{below a set limit of} 
%less than 
$340^\circ$ throughout the integration. Out of these Twotinos, 35 librate in the leading island, 21 in the trailing island, and the other 35 in the symmetric island.
%Out of these, 35 and 21 objects librate in the leading and trailing islands, respectively, while 35 librate in the symmetric island.

Figure \ref{fig:phitime} shows the 2:1 and 4:3 resonant arguments over \ccc{1.5 Myr} of four selected objects, representing the leading, trailing, and symmetric islands of the 2:1 MMR, as well as a non-resonant (circulating) object. The top panel shows the evolution of the 2:1 resonant argument, while the bottom panel corresponds to the evolution of the 4:3 resonant argument, as described in equation \ref{equ:phi43}. In the bottom panel of Fig. \ref{fig:phitime}, the leading object (blue curve) closely follows Pluto's resonant argument, as it has a near-zero libration amplitude in the 2:1 MMR. In contrast, the object on the trailing island (green curve) exhibits amplitude modulations throughout its evolution. The object in the symmetric island (red curve) shows secondary peaks, where the slower evolution will result in preferred longitudinal angles. In contrast, the non-resonant orbit (gray curve) circulates in both panels with a continuously increasing amplitude.

\ccc{To highlight the effect of Pluto on the evolution of Twotinos, we also show a red dashed line representing the evolution of \dpn{2015}{KX}{173} (the symmetric object) when turning off Pluto's influence. Pluto's effect seems negligible within 1 Myr, and statistically insignificant at 1.5 Myr, however after 4 Gyr, it will result in completely different orbits for some Twotinos. Although imperceptible in Fig. \ref{fig:phitime}, there are dashed lines for the other 3 objects, but they need approximately 10 Myr to show noticeable differences.}

%, which results from the definition of $\phi_{4:3}$, and the fact that $\phi_{2:1}$ is discontinuous while limited to the $0^\circ$ - $360^\circ$ range. 

%The behavior of the gray curve in the bottom panel of Figure \ref{fig:phitime} is due to the way we defined $\phi_{4:3}$, and the fact that $\phi_{2:1}$ is discontinuous while limited to the $0^\circ$ - $360^\circ$ range.

All objects in the symmetric island (as well as 5 objects in the asymmetric islands) have $\phi_{4:3}$ libration amplitudes larger than $360^{\circ}$; in fact, most of them have libration amplitudes larger than $720^{\circ}$. Although this could be interpreted as a form of circulation, this circulation is not the same as that observed in the non-resonant example. 

%In conventional three-body systems, once $360^\circ$ degrees have been passed, the same configuration is reached again. In our case, the influence of the resonances with Neptune prevents the system from returning to the same position. This kind of behavior of $\phi_{4:3}$ will cause certain angles to be preferred.

All remaining Twotinos show behaviors similar to the above four examples, 
%described by the linear combination of $\phi_{2:1}$ and $\phi_{3:2}$, 
although with secondary amplitude modulations due to varying frequencies between the 3:2 and 2:1 resonant arguments. The complete figure set is uploaded to a GitHub repository along with our classification for each of the 91 Twotinos (as well as the 61 circulating objects near the resonance) \footnote{\url{https://github.com/Sramrev/Pluto-Twotino-MMR.git}}. This confirms that a 2:1 MMR between a TNO and Neptune directly correlates with a 4:3 MMR between the same object and Pluto. 

%\footnote{Weaker resonant angles also librate in some of our 10 Myr simulations, such as the ones corresponding to equations 4D1.6 and 4D1.8 from \cite{murray99}. This is mainly because the libration of Pluto's argument of periapsis librates as well.}

% Although unconventional, in this work we assume that $\phi_{43}$ can span more than $360^\circ$. In conventional three-body systems, once $360^\circ$ degrees have passed, the same configuration is reached again. In this case, the influence of the resonances with Neptune allows the resonant angle to wind multiple times before retracing its path back again, coming back again to its starting position. This kind of behavior of $\phi_{43}$ will cause certain angles to be preferred in the same way that resonances with lower libration amplitudes do.

\section{Dynamical Significance of the 4:3 Resonance} \label{Sec:DynamicalSignificance}

\ccc{The presence of a MMR between two objects implies not only a geometrical conmensurability, but a dynamical enhanced interaction tied to the geometrical configuration. In general, this means we could expect: 
(A) a persistent $p\,$:$\,q$ ratio between the orbital periods,
(B) an enhanced gravitational interaction, either between the two objects, or, at least, from the more massive object affecting the lighter one, and
(C) that the gravitational perturbation carves a region of phase-space where the $p$:$q$ ratio is maintained and reinforced by the resonant interaction.}

\ccc{The Pluto-Twotino interaction described in the previous section fulfills two out of the three characteristics of the MMR interaction (A+B), since Pluto affects the evolution of Twotinos secularly \citepalias[as has been proved in][see e.g., Fig. 7]{Munoz26}; this effect is almost certainly due to the persistent 4:3 synchronous gravitational interactions between Pluto and Twotinos. Although Pluto's small mass is incapable of carving the Twotino phase-space (point C), they are nonetheless in interaction thanks to their mutual resonant relation with Neptune.}

%52222 \ccc{Thinking about a mean motion resonance between two objects, we would expect to observe three things: i) a persistent $p:q$ orbital period ratio; ii) a relevant gravitational interaction, either between 2 particles, or, at least, from the more massive one affecting the lighter one; and iii) that the gravitational interaction carves a region of phase-space where the p:q ratio is enhanced. While a purely geometrical effect would only fulfill the first one as the gravitational interaction would not be relevant and thus the heavier object would be incapable of carving a region in the phase-space with an enhanced gravitational interaction.}

%\ccc{In \citet{Munoz26} we can see the interaction between Pluto and the Twotinos, do not correspond to a geometrical effect, as there is a clear gravitational effect. i.e. the Pluto-Twotino interaction fulfills 2 out of the 3 requisites of a traditional resonance; less than a traditional resonance but more than a mere geometrical effect.}

%To further exemplify the significance of this interaction, ...

%%%%%%%%%%%%%%%%%%%%%%%%%%%%%%%%%%%%%%%%%%%%%%%%%%%%%%%%%%%%%%%%%%%%%%

%NOTA PARA EL PROXIMO ARTICULO!!!!!!!!!!!!!!
%*********************
%NO BORRAR
%Batigin, Morbidelli, 2013. ecuacion 3. El hamiltoniano se puede dividir en 6 S-N, S-P, S-T, N-P, N-T y P-T. Los primeros 3 son kepplerianos, el 4 y 5 son resonancias seguras y dadas por entendidas, y el 6 es la resonancia 43-PT

\ccc{To further exemplify the significance of the Pluto-Twotino MMR,} Figure \ref{fig:corot} shows the co-rotating orbits of Twotinos relative to Pluto, corrected via the mean longitude of the latter. Objects in the asymmetric islands, with libration amplitudes of $174^\circ$ for the leading Twotino (in blue) and $229^\circ$ for the trailing Twotino (in green), occupy the same longitudinal angles in the co-rotational frame at perihelion. The white curves superimposed on the colored curves represent the first 62,000 yr of each orbit, equivalent to the slower libration period of our example Twotinos.
%i.e., the one librating in the symmetric island (shown in red). 
%i.e. \dpn{2015}{KX}{173} (shown in red).
%This period is also shown by the vertical dashed lines in figure \ref{fig:phitime}. 
The confinement of the allowed angles in a co-rotational frame to a restricted range of values is characteristic of resonant dynamics. In the case of these Twotinos, such confinement allows them to avoid, or to regulate, the strength of gravitational perturbations from Pluto during conjunctions.

% Looking at the corresponding bottom panel, the leading object has six preferred angles grouped by pairs, where the distance between two peaks in the same pair is given by the amplitude of the resonant angle. This would also be the case for the trailing object, but each pair is broadened due to the amplitude modulations seen in figure \ref{fig:phitime}, resulting in each pair merging in a single major peak. 

The sample Twotino librating in the symmetric island is shown in red in Figure \ref{fig:corot}. Although libration amplitudes of the 4:3 resonant argument larger than $360^\circ$ technically permit $\phi_{4:3}$ to go around the entire circumference in the co-rotating frame, the nature of this particular resonance will result in some angles being preferred over others, i.e., the behavior of $\phi_{4:3}$ in an extended range prevents all angles from being repeated homogeneously. 

This behavior is further confirmed in the lower panels of Figure \ref{fig:corot}, which show the probability distribution of visited co-rotating angles for each of the selected Twotinos. In these panels, the leading and trailing objects have clear preferred angles. Among the 56 objects in the asymmetric islands, the ratios between favored and unfavored angles cover a 1.5 - 4 factor range. 
Librating objects with $360^\circ < \phi_{4:3} < 850^\circ$ show the same effect as particles with librations $\phi_{4:3}< 360^\circ$, although the peaks in the histograms are less pronounced; the sample symmetric object has the second highest signal among objects in the symmetric island, with a favored to unfavored angles ratio of 1.7; all other symmetric Twotinos have values between 1 and 1.9, with an average value of $\sim1.3$.
%average value of 1.32.

%Even when the libration amplitude of $\phi_{4:3}$ exceeds $360^\circ$, the resonant configuration in a co-rorational frame allows for the repetition of pulls between objects in a limited set of directions without being canceled by pulls from all other directions.
%; thus justifying our decision to consider them resonant despite their unusually large libration amplitude. 
This is in striking contrast to the circulating particles near the 2:1 MMR, one such example shown in gray in Figure \ref{fig:corot}, which, apart from covering the entire circumference in the co-rotating frame, shows no preferred angles in the probability density histogram. 

%These panels show peaks separated by $\sim60^\circ$ intervals (other objects have only 3 peaks), showing a behavior similar to the leading and trailing resonant particles, albeit as a diluted effect. 

% Objects in the symmetric island are not bound to the same pairing behavior, as the sweeping of the angle may go beyond $120^\circ$, the secondary peak may land at completely separate positions. In this case, the corresponding pairs are $\60^\circ$ apart, corresponding to the resonant angle stopping every $180^\circ$. The relative peak heights is 1:1 for the first 0.6 Myr, and then becomes 2:1 when the angle behavior shifts to stopping every $270^\circ$.

% The histogram max/min ratio mean for asymmetric islands is 2.75

\subsection{Comparison with Eris}\label{Sec:OtherCons}

%\section{Comparision with Eris}\label{Sec:OtherCons}

%Not only is Pluto's presence able to affect the behaviour of Twotinos, it is also capable of modifying the behaviour of Neptune: Neptune's mass is large enough that the orbital semi-major axis ratio is close enough, and the critical angle is in the sweet spot, so that Neptune’s presence dominates Pluto's orbit trapping it into the 3:2 MMR; what is sometimes forgotten is that Neptune is also trapped with Pluto and the details of Pluto orbit affect Neptune’s orbit \citep[e.g.][]{XXXXXX}. i.e. if one models the Solar System as the Sun, the four giant planets and a massless Pluto, Pluto is less well trapped than in a model where Plutos mass is included; this is not Neptune being able to trap Pluto better as Pluto’s mass will not affect the acceleration produced by the gravity of Neptune ass, a small contribution due to Pluto ``faintly'' trapping Neptune. 

One might think that Pluto affects Twotinos because its aphelion reaches Twotinos' semi-major axis. In this scenario, we would also expect Eris to affect Twotinos and even Plutinos since its orbit travels between 38 and 98 au; however, our studies have shown that Eris's effect is at least an order of magnitude smaller than Pluto's. This occurs even with Eris (plus Dysnomia) being $\sim14\%$ more massive than Pluto (plus Charon). 

In \citetalias{Munoz26} we studied Twotino leaking rates, and found that, over 4 Gyr, considering the four giant planets, without Pluto, the fraction of stable Twotinos is close to 47\%. If Eris is included as a massive perturber, the fraction of stable Twotinos decreases to 46\%. Instead, when Pluto was included as a massive perturber (without Eris), this fraction fell to 19\%. For Plutinos, the stable fractions correspond to 77\%, 77\%, and 36\%, in the same previous scenarios, respectively. This is likely because, while neither Eris nor Pluto can trap TNOs in their MMRs, Neptune will trap TNOs in Pluto's resonances, but not in Eris's. Once locked in, Pluto will slightly yet \ccc{steadily} affect resonant objects on secular time-scales, while Eris will not.

%%%%%%%%%%%%%%%%%%%%%%%%%%%%%%%%%%%%%%%%%%%%%%%%%%%%%%%%%%%%%%
\section{Conclusions}\label{Sec:Conclusions}

We have shown that a single resonant argument of the 4:3 MMR between Pluto and Twotinos librates. This means that such a resonance is relevant for the secular evolution of Twotinos, and this could be the origin of their increased instability in simulations when Pluto is present, compared to cases where only the four giant planets are included. The amplitude libration of the 4:3 MMR is conventional, i.e., below $360^{\circ}$ for most Twotinos librating in the asymmetric islands of the 2:1 MMR with Neptune. \ccc{For symmetric Twotinos, the libration amplitude of up to $850^{\circ}$ exceeds the typically considered hard limit of $360^{\circ}$; nonetheless,}
%is much larger than the $360^{\circ}$ limit typically considered in resonance studies; nonetheless, 
there is a limited set of angular values that are preferentially visited by those Twotinos in the co-rotating frame of Pluto. Such behavior results in the same resonant phenomena as for Twotinos in the asymmetric islands, although with a weakened effect. 

This result forces us to reconsider Pluto as an important component when studying the long-term evolution of the trans-Neptunian region, \ccc{specifically} for resonant populations such as Plutinos and Twotinos. With modern computers, the inclusion of Pluto represents a minimal effort, with almost imperceptible cost in integration time, so excluding it should not be considered a valid option. 

%%%%%%%%%%%%%%%%%%%%%%%%%%%%%%%%%%%%%%%%%%%%%%%%%%%%%%%%%%%%%%
\begin{acknowledgements}

\ccc{We thank the anonymous referee for the careful review that helped us to improve this letter.} This research was performed using services/resources provided by Grid UNAM, which is a collaborative effort driven by DGTIC and the research institutes of Astronomy, Nuclear Science and Atmosphere Science and Climate Change at UNAM. A.P.-V. acknowledges the DGAPA-PAPIIT grant IN112526. M.A.M.-G. acknowledges Universidad de Atacama for the DIUDA grant No. 88231R14.

\end{acknowledgements}

%%%%%%%%%%%%%%%%%%%%%%%%%%%%%%%%%%%%%%%%%%%%%%%%%%%%%%%%%%%%%%
% WARNING
% Please note that we have included the references below in
% order to compile the document, but we ask you to:
%
% - use BibTeX with the regular commands:
   \bibliographystyle{aa} % style aa.bst
   \bibliography{pluto43} % your references Yourfile.bib
% - join the .bib files when you upload your source files
%%%%%%%%%%%%%%%%%%%%%%%%%%%%%%%%%%%%%%%%%%%%%%%%%%%%%%%%%%%%%%

%\begin{thebibliography}{}

%  \bibitem[Baker(1966)]{baker} Baker, N. 1966,
%      in Stellar Evolution,
%      ed.\ R. F. Stein,\& A. G. W. Cameron
%      (Plenum, New York) 333

%\end{thebibliography}

% %%%%%%%%%%%%%%%%%%%%%%%%%%%%%%%%%%%%%%%%%%%%%%%%%%%%%%%%%%%%%%
% Example below of non-structurated natbib references  
% To use the v8.3 macros with this form of composition of bibliography,
% the option "bibyear" should be added to the command line
% "\documentclass[bibyear]{aa}".
% %%%%%%%%%%%%%%%%%%%%%%%%%%%%%%%%%%%%%%%%%%%%%%%%%%%%%%%%%%%%%%

%%%%%%%%%%%%%%%%%%%%%%%%%%%%%%%%%%%%%%%%%%%%%%%%%%%%%%%%%%%%%%%
% Appendices must be placed after   \end{thebibliography}
% They will be placed automatically on a new page.
%%%%%%%%%%%%%%%%%%%%%%%%%%%%%%%%%%%%%%%%%%%%%%%%%%%%%%%%%%%%%%%
\begin{appendix}
%%%%%%%%%%%%%%%%%%%%%%%%%%%%%%%%%%%%%%%%%%%%%%%%%%%%%%%%%%%%%%%
% In the PDF output, floats should be placed
% under their own appendix, not before the title, nor after the
% title of the next appendix.

% In short appendices, onecolumn floats (\figure*
% or \table*) will generate a blank page.
% To prevent this behaviour, a few examples are provided here. 

% In case you have a lot of floating objects for little text and the 
% LaTeX engine moves the floats away from their context, the command
% \FloatBarrier of the “placeins” package will empty the
% float buffer and place all stored floats in the continuity.

% If you still encounter problems with wide floats placement,
% just use the onecolumn environment throughout the appendices.
%%%%%%%%%%%%%%%%%%%%%%%%%%%%%%%%%%%%%%%%%%%%%%%%%%%%%%%%%%%%%%%

%____________________________________________________________
%       Wide floats at the start of an appendix: first method
%-------------------------------------------------------------
% To prevent a blank page after the start of an appendix:
% - Switch to one \onecolumn first
% - Declare the section title
% - Declare the onecolumn float with the parameter [ht!]
% - Revert to \twocolumn at the end of the section

\end{appendix}
\end{document}